\documentclass[portuguese,
  manuscript=article,  
  layout=publish,
  year=2023,
  volume=3,
]{jmfis}

\usepackage[brazilian]{babel} 
\usepackage{amsmath}
\usepackage{dsfont}
\usepackage{bm}
\usepackage{cite}
\usepackage{graphicx}
\hypersetup{pdfborder = 0 0 0}

\doi{{ {\color{blue}\href{https://doi.org/10.59396/29651964JMFis.3.11.2023}{10.59396/29651964JMFis.3.11.2023}}}}

\received {11/07/2023}
\accepted {03/08/2023}
\published{08/08/2023}

\newcommand{\orcidlink}[1]{\href{https://orcid.org/#1}{\includegraphics[width=10pt]{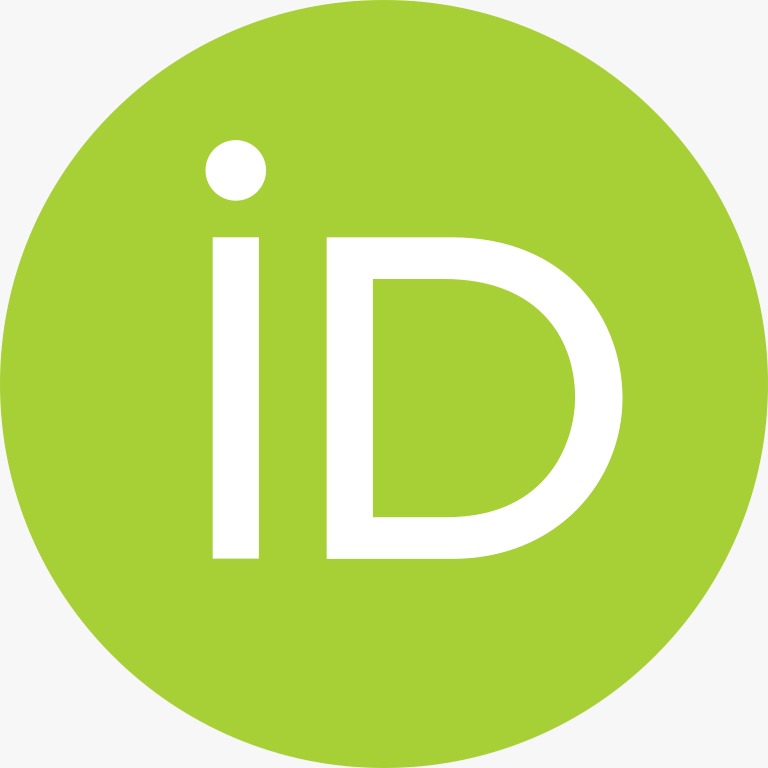}}}

\title{Phase Transitions in a Network with Assortative Mixing}

\author{R. A. Dumer \orcidlink{0000-0002-9032-111X}}
\email{rafaeldumer@fisica.ufmt.br}
\affiliation{Instituto de Física - Universidade Federal de Mato Grosso, 78060-900, Cuiabá, Mato Grosso, Brasil.}

\author{M. Godoy \orcidlink{0000-0001-9122-6061}}
\email{mgodoy@fisica.ufmt.br}
\affiliation{Instituto de Física - Universidade Federal de Mato Grosso, 78060-900, Cuiabá, Mato Grosso, Brasil.}

\palavrachave{Modelo de Ising; Comportamento crítico; Mistura assostativa}

\begin{document}
\begin{resumo}
 Neste trabalho, utilizamos o modelo de Ising para identificar transições de fase em um sistema magnético, onde a distribuição de graus da rede obedece uma lei de potência e as conexões são feitas por uma mistura assortativa. No modelo de Ising os spins assumem apenas dois valores, $\sigma=\pm1$, e interagem através de um acoplamento ferromagnético $J$. A rede tem quatro parâmetros variáveis:  o expoente da distribuição de graus $\alpha$, o grau mínimo $k_0$, o grau máximo $k_m$, e $p_{r}$ que pode representar o grau de assortatividade ou dissortatividade da rede. Com objetivo de estudar o efeito das correlações dos graus da rede no comportamento crítico do sistema, fixamos $k_0=4$, $k_m=10$, e $\alpha=1$, e variamos $p_r$ de modo a obter uma mistura assortativa das arestas. Como resultado calculamos os pontos de transição de fase do sistema, e os expoentes críticos relacionados a magnetização $\beta$, susceptibilidade magnética $\gamma$, e comprimento de correlação $\nu$.
\end{resumo}

\vspace{0.5cm}

\begin{abstract}
In this work, we employed the Ising model to identify phase transitions in a magnetic system where the degree distribution of the network follows a power-law and the connections are assortatively mixed. In the Ising model, the spins assume only two values, $\sigma = \pm 1$, and interact through ferromagnetic coupling $J$. The network is characterized by four variable parameters: $\alpha$ denotes the degree distribution exponent, the minimum degree $k_0$, the maximum degree $k_m$, and the $p_r$ represents the assortativity or disassortativity of the network. To investigate the effect of degree correlations on the critical behavior of the system, we fix $k_0=4$, $k_m=10$, and $\alpha=1$, and vary $p_r$ to obtain an assortative mixing of edges. As result, we have calculated the phase transition points of the system, and the critical exponents related to magnetization $\beta$, magnetic susceptibility $\gamma$, and the correlation length $\nu$.
\end{abstract}

\section{Introduction}
Understanding nature and society as networks, composed of vertices and edges, allows us to observe patterns that were previously mere speculations before quantitative analysis. One of these patterns is the level of correlation between the degrees of a network, where the degree is understood as the number of edges connected to a vertex. This correlation can be identified in two ways \cite{1}: (i) vertices with higher degrees tend to connect with vertices of higher degrees; (ii) vertices with higher degrees tend to connect with vertices of lower degrees. In the first case, networks with this characteristic are said to be assortative, while in the second case, they are referred to as dissortative networks. The most common examples of assortative networks are those related to collaborations in scientific work \cite{2}, or actors in films \cite{3}, whereas dissortativity is more commonly found in biological networks \cite{3,4}.

In this context, the study of correlated networks becomes important in terms of their structural aspects and consequences when implemented in systems highly dependent on the structure of the network, such as magnetic systems. In addition to correlated networks, the study of networks with a degree distribution following a power law also becomes interesting, as advances in the understanding of the dynamics and topological stability of networks reveal that networks, even from very distinct origins, self-organize with such a distribution \cite{5,6}. This means that regardless of the system and its contributors, the probability $P(k)$ that a vertex interacts with $k$ other vertices in the network decays as $P(k)\sim k^{-\alpha}$, where $\alpha$ is the exponent that depends on the system in question.

Recent work on magnetic systems has been conducted using networks with a degree distribution in the form of a power law, limiting both the maximum and minimum degrees of the network \cite{7,8,9}. In these studies, the network exhibited a small degree of assortativity, which may have influenced the phase transition point of the system, but no study has yet been conducted with the aim of identifying the influence of degree correlations on the critical behavior of magnetic systems. Thus, the objective of the present work is to use the Xulvi-Brunet and Sokolov algorithm \cite{10} to modify the degree of correlation of the network used in previous studies \cite{7,8,9}, and to investigate the influence of the assortativity of the network on the critical behavior of the system, both in terms of the phase transition point and the static critical exponents of the system.

\section{Model and Methods}

Initially, we created a network with a minimum degree of $k_0=4$, a maximum degree of $k_m=10$, and a degree distribution exponent of $\alpha=1$. These values were fixed to simplify the study, as we are primarily interested in the behavior of the system with different degrees of correlation, and these parameters were chosen because they yielded the most accurate results in previous research \cite{7,8,9}. To quantify the degree of correlation in the network, we used the Pearson correlation coefficient for networks with arbitrary degree distributions, as proposed by Newman \cite{1}:

\begin{equation}\label{eq1}
r=\frac{M^{-1}\sum_{i}u_{i}v_{i}-\left[M^{-1}\sum_{i}\frac{1}{2}\left(u_{i}-v_{i}\right)\right]^{2}}{M^{-1}\sum_{i}\frac{1}{2}\left(u_{i}^{2}+v_{i}^{2}\right)-\left[M^{-1}\sum_{i}\frac{1}{2}\left(u_{i}+v_{i}\right)\right]^{2}},
\end{equation}

where $M$ is the number of edges in the network, with $u_i$ and $v_i$ being the degrees of the ends of edge $i$. In this coefficient, $r>0$ indicates that the network is assortative, $r<0$ means the network is dissortative, while $r=0$ is characteristic of a neutral network. Naturally, our initial network already has $r\approx0.22$, indicating that it is assortative.

To vary the degree of correlation, we used the Xulvi-Brunet and Sokolov algorithm \cite{10}. In this algorithm, given the initial network, at each step we randomly select two edges and arrange their degrees in descending order, assuming $k_a \geq k_b \geq k_c \geq k_d$. We then define a probability $p_r$ related to the desired degree of correlation and randomly draw a number $\xi$ between zero and one. If, we want an assortative network and $\xi \leq p_r$, the selected edges are rewritten so that one will have degrees $k_a$ and $k_b$, while the other will have degrees $k_c$ and $k_d$. However, if we are interested in a dissortative network and $\xi \leq p_r$, the edges are rewritten so that one will have degrees $k_a$ and $k_d$, while the other will have degrees $k_b$ and $k_c$. In both cases, if $\xi > p_r$, the edges are rewritten randomly. Given the structure of the initial network, we used $t=6\times 10^5$ steps to reach network stability for the defined $p_r$, across all tested network sizes.

In order to implement this network in a magnetic system, we used the Ising model with a Hamiltonian describing the interaction between spins in the network as follows:

\begin{equation}\label{eq2}
{\cal H}=-J\sum_{\langle i,j\rangle}\sigma_{i}\sigma_{j},
\end{equation}

where we define the ferromagnetic coupling constant $J=1$, $\sigma_i=\pm1$, and the summation is performed over all pairs of spins connected in the network.

Now, to obtain the results, we used Monte Carlo simulations with the Metropolis algorithm, discarding the first $10^5$ Monte Carlo steps (MCS) and performing thermal averaging over $9\times10^5$ MCS with 10 independent samples. The thermodynamic quantities calculated were: the magnetization per spin $m_{N}=\left\langle m_{N}^1 \right\rangle$, the susceptibility $\chi_{N}$, and the Binder cumulant $U_{N}$:

\begin{equation}\label{eq3}
\left\langle m_{N}^n \right\rangle=\left[\left(\frac{1}{N}\sum_{i=1}^{N}\sigma_{i}\right)^n\right] 
\end{equation}

\begin{equation}\label{eq4}
\chi_{N}=\frac{N}{k_{B}T}\left( \left\langle m_{N}^{2}\right\rangle -\left\langle m_{N}^1\right\rangle ^{2} \right),
\end{equation}

\begin{equation}\label{eq5}
U_{N}=1-\frac{\left\langle m_{N}^{4}\right\rangle }{3\left\langle m_{N}^{2}\right\rangle ^{2}},
\end{equation}

where [...] denotes the average over the MCS. These quantities obey the following finite-size scaling relations for complex networks in the vicinity of the critical point:

\begin{equation}\label{eq6}
m_{N}=N^{-\beta/\nu}m_{0}(N^{1/\nu}\epsilon),
\end{equation}

\begin{equation}\label{eq7}
\chi_{N}=N^{\gamma/\nu}\\chi_{0}(N^{1/\nu}\epsilon),
\end{equation}

\begin{equation}\label{eq8}
U_{N}=U_{0}(N^{1/\nu}\epsilon),
\end{equation}

where $\epsilon=(T-T_{c})/T_{c}$, $m_{0}$, $\chi_{0}$, and $U_{0}$ are scaling functions, and $\beta$, $\gamma$, and $\nu$ are the critical exponents of the thermodynamic quantities.

\begin{figure}
\begin{centering}
\includegraphics[scale=1.3]{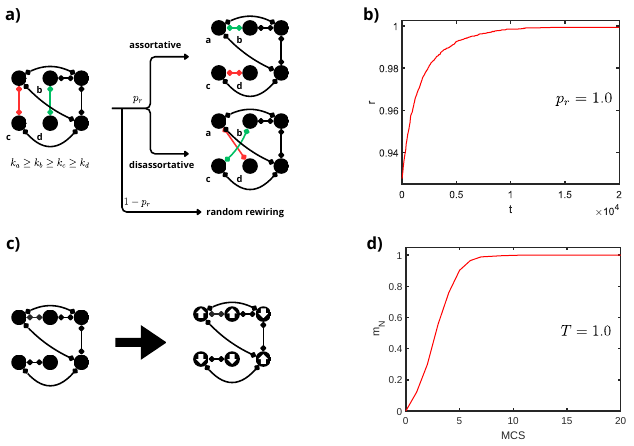}
\par\end{centering}
\caption{{\footnotesize{}In (a), we present a schematic representation of a step in the Xulvi-Brunet and Sokolov algorithm, as also mentioned in the methods section of this work, for a specific value of $p_r$. At each step of this algorithm, the network is modified, and $r$ is calculated as specified in Eq. (\ref{eq1}). This process is repeated until the network reaches a stationary state, where $r$ fluctuates around a certain value, as shown in (b) for $p_r=1.0$, with $r$ plotted as a function of the $t$-steps of the algorithm. Once the stationary state is reached, the network is saved to an external file for later use in simulating the Ising model. In this context, each vertex is then interpreted as a spin, as depicted in (c), and from this point, we perform Monte Carlo simulations using the Metropolis algorithm until the system reaches another stationary state. This time, the focus is on analyzing the system's magnetization, as illustrated in (d) for $T=1.0$.}}
\label{fig11}
\end{figure}

To better understand how the results will be obtained, we present the scheme in Fig.\ref{fig11}, which illustrates the creation of the network with a specified $p_r$ and its implementation in the Metropolis algorithm for calculating the magnetization of the system.

\section{Results}

\begin{figure}
\begin{centering}
\includegraphics[scale=0.4]{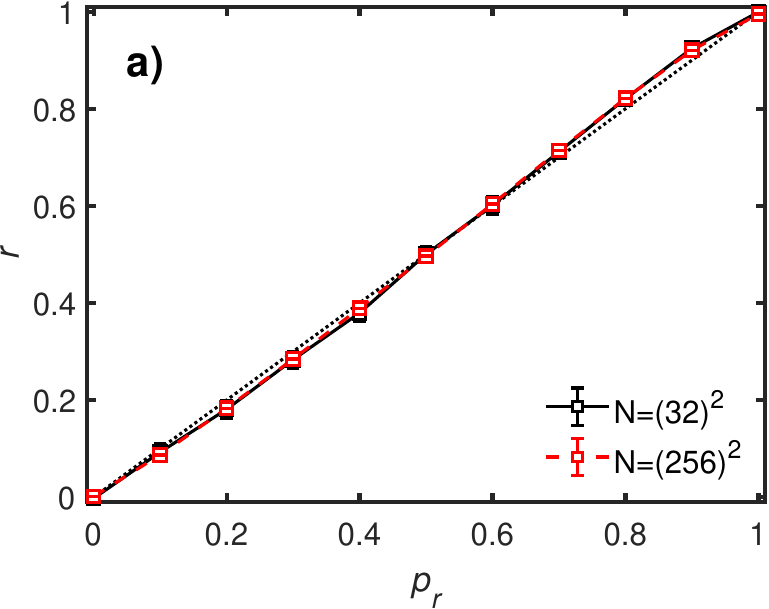}\hspace{0.5cm}\includegraphics[scale=0.4]{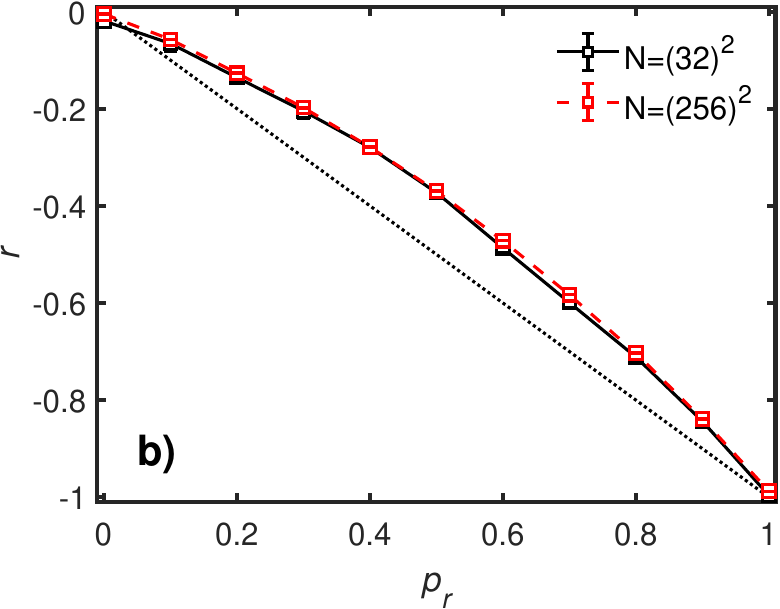}
\par\end{centering}
\caption{{\footnotesize{}(a) The degree correlation coefficient of the network as a function of $p_r$ for the case of an assortative mixing of edges is shown. In (b), this correlation is presented for the case of a dissortative mixing of edges. The curves are shown for two network sizes, $N=(32)^2$ and $N=(256)^2$, indicating that the correlation is independent of the system size. The dotted line is merely a reference for the linear behavior of $r$ as a function of $p_r$.}}
\label{fig1}
\end{figure}

In a fully assortative network, $r=1$, the vertices of the same degree should only connect to each other, but this will not occur in our case, as we initially do not have a disconnected network. The same applies to a fully dissortative network, $r=-1$, since the vertices with higher degrees should only connect to vertices with lower degrees, which would disrupt the degree distribution. Therefore, in this first step, we investigated the behavior of $r$ as a function of $p_r$ to identify the possible degrees of correlation that we can achieve with the network. In Fig. \ref{fig1} (a), we can observe this behavior for an assortative network, while in Fig. \ref{fig1} (b), our objective was to achieve a dissortative network. In both cases, we were able to reach approximate values of $r$ for a given $p_r$; however, this behavior deviates from linearity when we aim for $r<0$.

\begin{figure}
\begin{centering}
\includegraphics[scale=0.4]{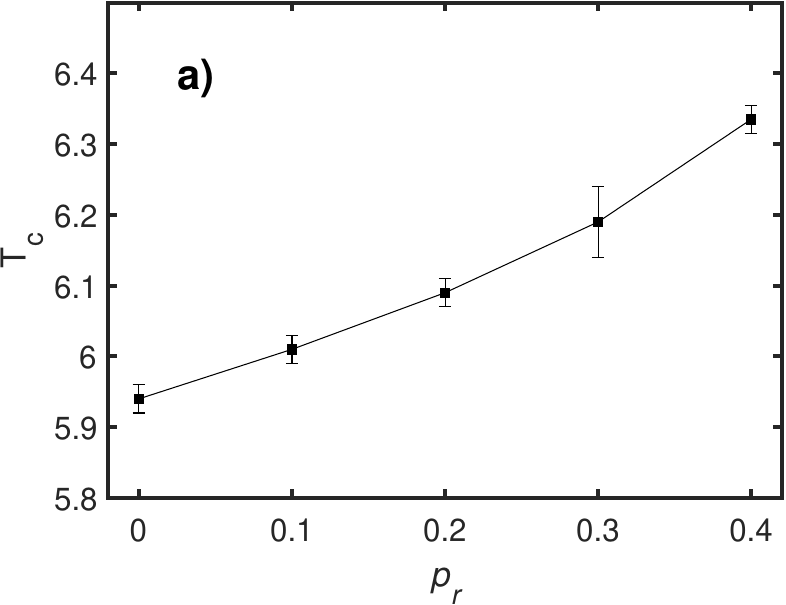}\hspace{0.5cm}\includegraphics[scale=0.4]{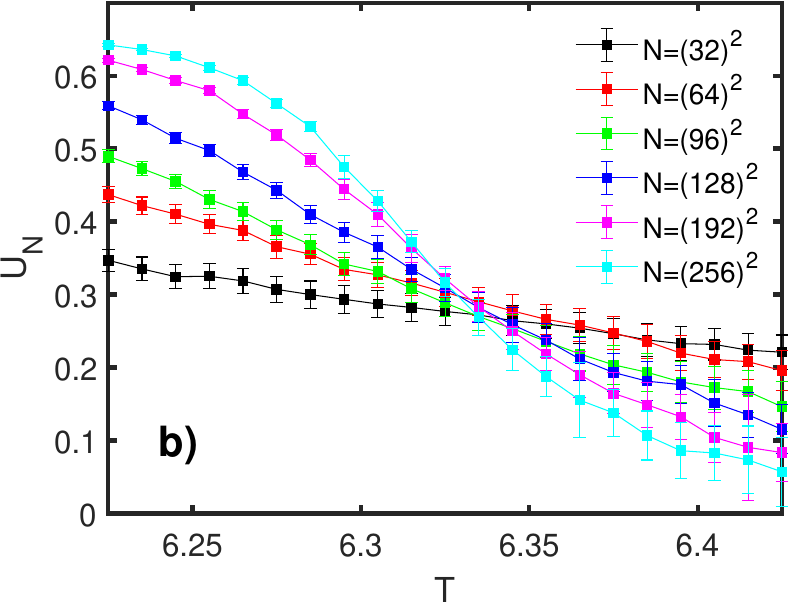}
\par\end{centering}
\caption{{\footnotesize{}(a) Phase diagram for different degrees of assortativity $p_r$, where below the curve we have the ferromagnetic phase and above the curve is the paramagnetic phase. In (b), we have an example of the estimation of $T_c$ with the $U_N$ curves for different network sizes, showing the phase transition point for $p_r=0.4$ at $T_c=6.34\pm0.02$.}}
\label{fig2}
\end{figure}

\begin{figure}
\begin{centering}
\includegraphics[scale=0.4]{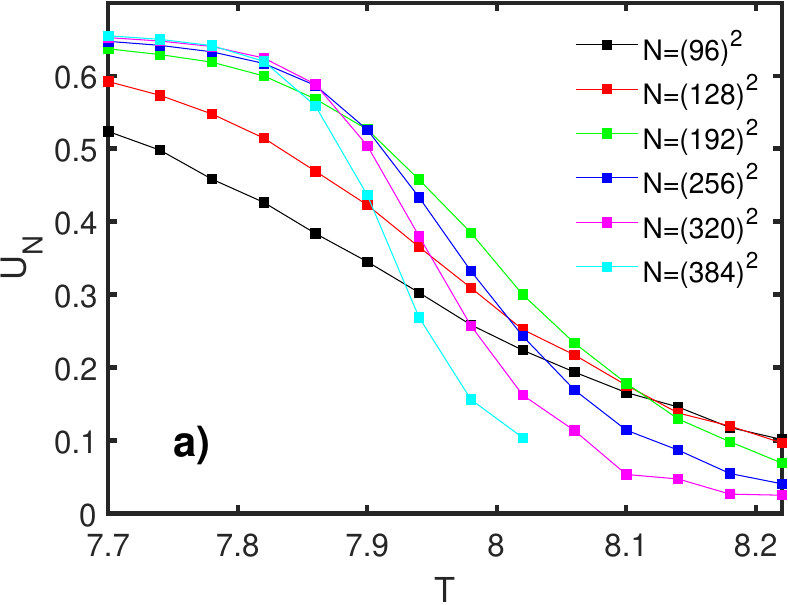}\hspace{0.5cm}\includegraphics[scale=0.4]{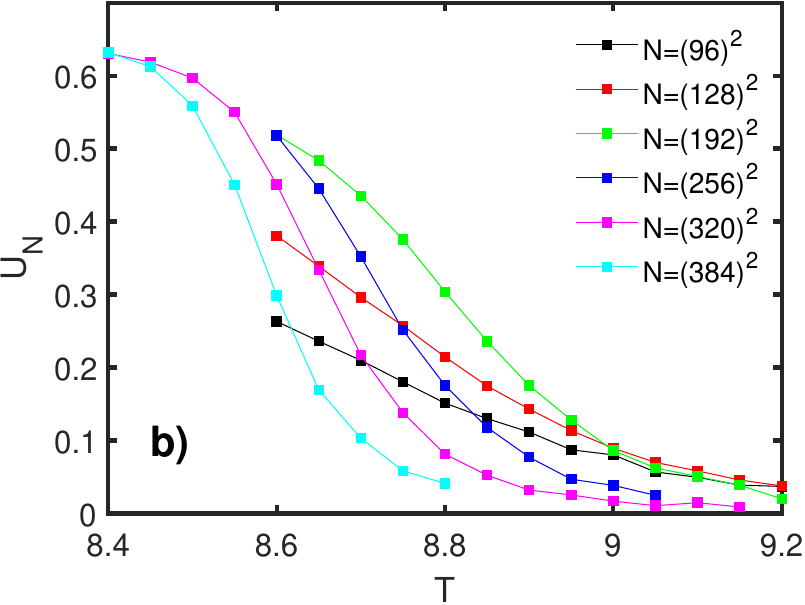}
\par\end{centering}
\caption{{\footnotesize{}Behavior of $U_N$ for different network sizes with $p_r=0.9$ in (a) and $p_r=1.0$ in (b).}}
\label{fig3}
\end{figure}

In studying the critical behavior of the system, we identified the phase transition points from the ferromagnetic (F) phase, at low temperatures $T$, to the paramagnetic (P) phase as $T$ increases. To obtain theses points, we utilized the crossing of the $U_N$ curves with different network sizes $N$, since at the phase transition point these curves should be independent of the system size. As a result, only for $p_r\le 0.4$, we able to obtain a well-defined critical point, as shown in the diagram in Fig. \ref{fig2} (a) and the example of the crossing of the $U_N$ curves in Fig. \ref{fig2} (b). Above $p_r=0.4$, as we increase the system size, the curves begin to shift towards lower temperatures, preventing a single crossing point for the curves. Consequently, we identified a crossover behavior, as larger network sizes correspond to lower phase transition temperatures. This shift in the $U_N$ curves can be seen in Fig. \ref{fig3} (a) for $p_r=0.9$ and in Fig. \ref{fig3} (b) for $p_r=1.0$.

\begin{figure}
\begin{centering}
\includegraphics[scale=0.4]{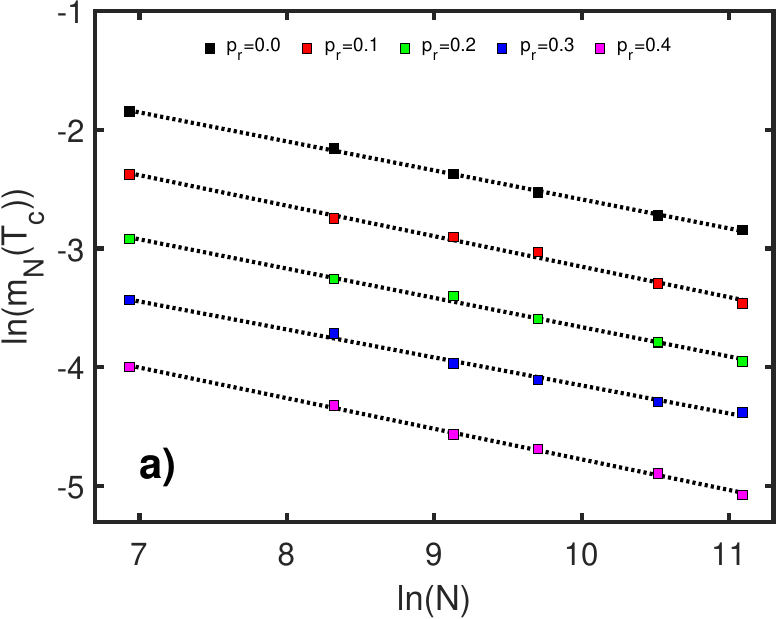}\hspace{0.5cm}\includegraphics[scale=0.4]{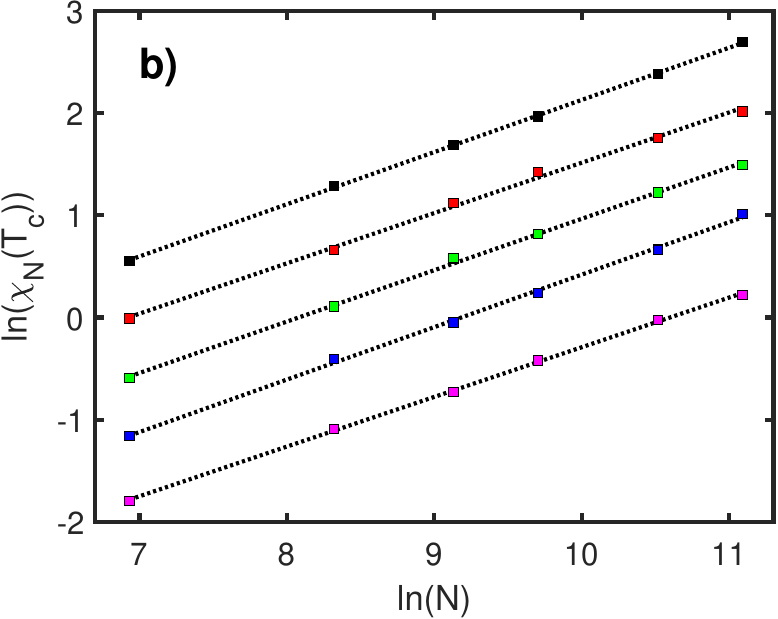}
\par\end{centering}
\vspace{0.1cm}
\begin{centering}
\includegraphics[scale=0.4]{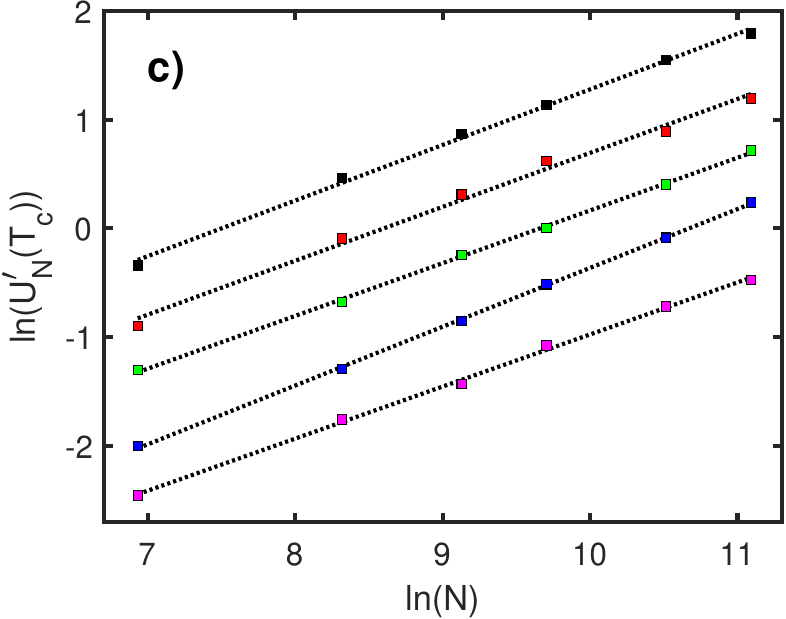}\hspace{0.5cm}\includegraphics[scale=0.4]{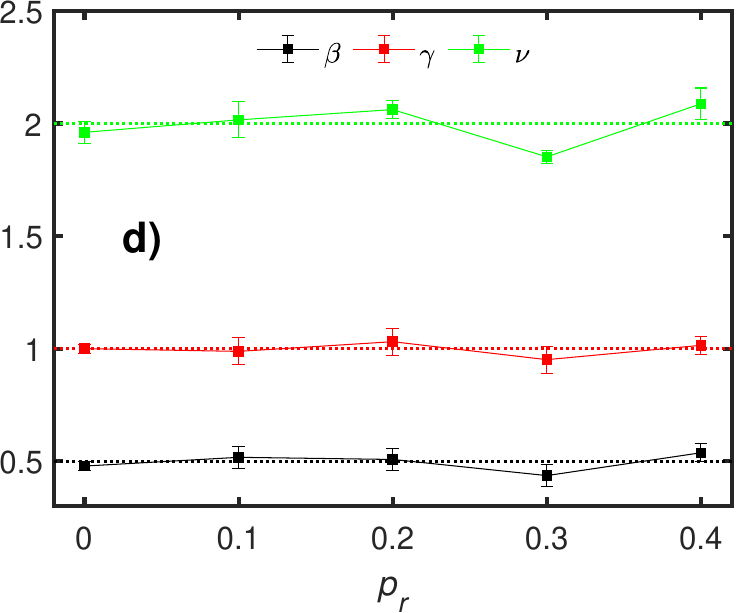}
\par\end{centering}
\caption{{\footnotesize{}In (a), (b), and (c) log-log plots for $m_N$, $\chi_N$, and the derivative of the Binder cumulant $U_N^{\prime}$, respectively, for different values of $p_r$ indicated in (a). In (d), we present the values of the critical exponents obtained from the slope of the linear fit of the curves in (a), (b), and (c). The dotted lines in (d) denote the critical exponents of the mean-field approximation: $\beta=0.5$, $\gamma=1.0$, and $\nu=2.0$.
}}
\label{fig4}
\end{figure}

Having identified the phase transition points, we can then calculate the critical exponents of the system. For this, we utilize the scaling relations from Eqs. (\ref{eq6}), (\ref{eq7}), and (\ref{eq8}). As these relations are valid in the vicinity of the critical point, we take the value of the thermodynamic quantity in this region as a function of the network size. Thus, in the graph with logarithmic scale axes, the slope of the resulting curves provides us with the appropriate ratios between the critical exponents.

In Fig. \ref{fig4} (a), the curves of magnetization based on the scaling relation from Eq. (\ref{eq6}) provide us with a slope that represents the ratio $-\beta/\nu$. Now, in Fig. \ref{fig4} (b), we have the curves for magnetic susceptibility in the vicinity of the critical point and which returns the ratio $\gamma/\nu$ as the slope, based on Eq. (\ref{eq7}). However, with only these two ratios, we can not separately identify the critical exponents. Therefore, to achieve this, we use the curves of the derivative of the Binder cumulant, which yield the ratio $1/\nu$, based on Eq. (\ref{eq8}). The curves for the derivative of the Binder cumulant can be seen in Fig. \ref{fig4} (c). With this, we can identify the exponents $\beta$, $\gamma$, and $\nu$ individually, and these are compiled in Fig. \ref{fig4} (d) as a function of $p_r$. In this figure, we observe that the exponents do not diverge from one another within the respective statistical errors and that they are compatible with the critical exponents of mean-field theory for complex networks, which are represented by the dotted lines in Fig. \ref{fig4} (d).

\section{Conclusions}
In this work, we have studied the Ising model on a network with a degree distribution following a power law, $P(k) \sim k^{-\alpha}$, with fixed values of minimum degree $k_0 = 4$, maximum degree $k_m = 10$, and exponent $\alpha = 1$. We altered the correlation degree of the network and examined its effects on the critical behavior of the system in the case of assortative mixture networks. The degree of correlation is directly related to the parameter $p_r$, and we observed a well-defined critical temperature for the system only for $p_r \leq 4$. Above this value, as the system size increases, the observed transition temperature decreases. As $p_r$ increases, we note that the critical temperature of the system also increases, corroborating previous results for networks with non-zero degree correlation \cite{7,8,9}. Regarding the critical exponents, we found that they are not affected by the different degrees of assortativity of the network, and as expected, they belong to the universality class of the mean field approximation.

\begin{acknowledgement}
We would like to thank the National Council for Scientific and Technological Development (Conselho Nacional de Desenvolvimento Científico e Tecnológico - CNPq), by the PhD scholarship granted for R. A. Dumer under the process number 140141/2024-3.
 \end{acknowledgement}


\end{document}